

\input epsf.tex

\font\rmu=cmr10 scaled\magstephalf
\font\bfu=cmbx10 scaled\magstephalf

\font\it=cmti10 scaled \magstephalf

\rmu

\font\rmus=cmr8
\font\rmuss=cmr6
\font\mait=cmmi10 scaled\magstephalf
\font\maits=cmmi7 scaled\magstephalf
\font\maitss=cmmi7
\font\msyb=cmsy10 scaled\magstephalf
\font\msybs=cmsy8 scaled\magstephalf
\font\msybss=cmsy7
\font\bfus=cmbx7 scaled\magstephalf
\font\bfuss=cmbx7
\font\cmeq=cmex10 scaled\magstephalf

\textfont0=\rmu
\scriptfont0=\rmus
\scriptscriptfont0=\rmuss

\textfont1=\mait
\scriptfont1=\maits
\scriptscriptfont1=\maitss

\textfont2=\msyb
\scriptfont2=\msybs
\scriptscriptfont2=\msybss

\textfont3=\cmeq
\scriptfont3=\cmeq
\scriptscriptfont3=\cmeq

\newfam\bmufam  \textfont\bmufam=\bfu
      \scriptfont\bmufam=\bfus \scriptscriptfont\bmufam=\bfuss

\hsize=15.5cm
\vsize=21cm
\baselineskip=16pt   
\parskip=12pt plus  2pt minus 2pt

\def\a{\alpha}
\def\b{\beta}
\def\d{\delta}
\def\e{\epsilon}

\def\g{\gamma}

\def\semi{\bigcirc\kern-1em{s}\;}

\def\del{\partial}
\def\ni{\noindent}

\def\one{{\mathchoice {\rm 1\mskip-4mu l} {\rm 1\mskip-4mu l}
{\rm 1\mskip-4.5mu l} {\rm 1\mskip-5mu l}}}
\def\Q{{\mathchoice
{\setbox0=\hbox{$\displaystyle\rm Q$}\hbox{\raise 0.15\ht0\hbox to0pt
{\kern0.4\wd0\vrule height0.8\ht0\hss}\box0}}
{\setbox0=\hbox{$\textstyle\rm Q$}\hbox{\raise 0.15\ht0\hbox to0pt
{\kern0.4\wd0\vrule height0.8\ht0\hss}\box0}}
{\setbox0=\hbox{$\scriptstyle\rm Q$}\hbox{\raise 0.15\ht0\hbox to0pt
{\kern0.4\wd0\vrule height0.7\ht0\hss}\box0}}
{\setbox0=\hbox{$\scriptscriptstyle\rm Q$}\hbox{\raise 0.15\ht0\hbox to0pt
{\kern0.4\wd0\vrule height0.7\ht0\hss}\box0}}}}
\def\C{{\mathchoice
{\setbox0=\hbox{$\displaystyle\rm C$}\hbox{\hbox to0pt
{\kern0.4\wd0\vrule height0.9\ht0\hss}\box0}}
{\setbox0=\hbox{$\textstyle\rm C$}\hbox{\hbox to0pt
{\kern0.4\wd0\vrule height0.9\ht0\hss}\box0}}
{\setbox0=\hbox{$\scriptstyle\rm C$}\hbox{\hbox to0pt
{\kern0.4\wd0\vrule height0.9\ht0\hss}\box0}}
{\setbox0=\hbox{$\scriptscriptstyle\rm C$}\hbox{\hbox to0pt
{\kern0.4\wd0\vrule height0.9\ht0\hss}\box0}}}}

\font\fivesans=cmss10 at 4.61pt
\font\sevensans=cmss10 at 6.81pt
\font\tensans=cmss10
\newfam\sansfam
\textfont\sansfam=\tensans\scriptfont\sansfam=\sevensans\scriptscriptfont
\sansfam=\fivesans
\def\sans{\fam\sansfam\tensans}
\def\Z{{\mathchoice
{\hbox{$\sans\textstyle Z\kern-0.4em Z$}}
{\hbox{$\sans\textstyle Z\kern-0.4em Z$}}
{\hbox{$\sans\scriptstyle Z\kern-0.3em Z$}}
{\hbox{$\sans\scriptscriptstyle Z\kern-0.2em Z$}}}}

\newcount\foot
\foot=1
\def\note#1{\footnote{${}^{\number\foot}$}{\ftn #1}\advance\foot by 1}

\def\frac#1#2{{#1\over #2}}
\def\text#1{\quad{\hbox{#1}}\quad}

\font\ch=cmbx12 scaled\magstephalf
\font\ftn=cmr8 scaled\magstephalf

\font\it=cmti10 scaled\magstephalf

\font\titch=cmbx12 scaled\magstep2
\font\titname=cmr10 scaled\magstep2
\font\titit=cmti10 scaled\magstep1
\font\titbf=cmbx10 scaled\magstep2

\nopagenumbers

\line{\hfil DFF 235/11/95}
\line{\hfil November 3, 1995}
\vskip2.2cm
\centerline{\titch SPECTRUM OF THE VOLUME OPERATOR }
\vskip.5cm
\centerline{\titch IN QUANTUM GRAVITY}
\vskip1.7cm
\centerline{\titname R. Loll\note{Supported by the European Human
Capital and Mobility program ``Constrained Dynamical Systems"}}
\vskip.5cm
\centerline{\titit Sezione INFN di Firenze}
\vskip.2cm
\centerline{\titit Largo E. Fermi 2}
\vskip.2cm
\centerline{\titit I-50125 Firenze, Italy}
\vskip.3cm
\centerline{and}
\vskip.3cm
\centerline{\titit Max-Planck-Institut f\"ur Gravitationsphysik}
\vskip.2cm
\centerline{\titit Schlaatzweg 1}
\vskip.2cm
\centerline{\titit D-14473 Potsdam, Germany}

\vskip2.0cm
\centerline{\titbf Abstract}
\vskip0.7cm

The volume operator is an important kinematical quantity in the
non-perturbative approach to four-dimensional quantum gravity in
the connection formulation. We give a general algorithm for computing
its spectrum when acting on four-valent spin network states, evaluate
some of the eigenvalue formulae explicitly, and discuss
the role played by the Mandelstam constraints.

\vfill\eject
\footline={\hss\tenrm\folio\hss}
\pageno=1


\line{\ch 1 Introduction\hfil}

The volume operator has emerged as an important
quantity in the kinematics of 3+1-dimensional quantum gravity
in the loop representation. It is the quantum analogue of the
classical volume function, measuring the volume of three-dimensional
spatial regions.
Although not an observable of the pure gravity theory (in the
sense of commuting with the gravitational Hamiltonian), it has
an immediate physical interpretation, and becomes a genuine observable
if the spatial regions are defined intrinsically by
additional matter fields,
for example, the constant-value surfaces of a scalar field variable.
Moreover, the volume is a relatively simple function of the
dreibein variable $E$, of the canonically conjugate pair $(A,E)$,
where $A\in\cal A$ is the (complex) $SU(2)$-Ashtekar connection on
the three-space $\Sigma$.

The significance of the volume operator for the quantum theory derives
from the fact that its eigenfunctions span the kinematical quantum
state space of functions on ${\cal A}/{\cal G}$, the space of
connections modulo gauge transformations, and that these functions are
known, at least in principle [1]. They are the so-called spin network
states, which can be expressed as appropriately (anti-)symmetrized
linear combinations of certain Wilson loop functions (traces of
holonomies of the connection variable $A$). This, as well as the
spectral discreteness of the volume operator, was first pointed out
in [2].

An obvious task at this stage is the actual computation of its
spectrum. We will describe in the following a general algorithm
for computing the volume spectrum on
four-valent spin network states, that is,
spin networks made up of Wilson loop states with no more than
four line segments meeting at each loop intersection.
The eigenvalues of several classes of such states will be
computed explicitly. We have shown
previously [3] that the spectrum in the three-valent case is
identically zero, thereby correcting a computational error in [2],
where a non-vanishing spectrum was derived. (Because of the algebraic
structure of the volume operator it is clear a priori that loop
states with valence less than three are annihilated.)

Our calculations will take place in a lattice-regularized framework
which we have been advocating elsewhere as an appropriate tool for
approximating the quantum Hamiltonian dynamics of gravity in the
loop approach [4]. However, to our understanding the form of the
lattice operator coincides with that of the continuum formulations [5,6]
(when restricted to the subset of states that can be realized on a
three-dimensional cubic lattice), and therefore our results are
equally valid in those cases. Since the discussion of the volume
operator is largely insensitive to the signature of space-time and the
particular form of the Hamiltonian, we will for simplicity work within
the real, $SU(2)$-setting. We only remind the reader that there
is a version of Ashtekar's gravity based on a real canonical variable
pair $(A,E)$, as recently discussed in [7] (where, however, the
Hamiltonian assumes a more complicated form than in the complex
formulation).

As part of the spectrum calculation, we will have to address the
issue of over-completeness of the spin network states, which is
a familiar feature of complete sets of gauge-invariant functions
on a space $\cal A$ of gauge potentials. We will argue that it is
necessary to eliminate this over-completeness in order to derive
the correct spectrum for the volume operator.

In the next section, we will recall the construction of the classical
volume function and the discretized quantum volume operator on the
lattice. In Sec.3, we derive a general expression for the matrix
elements of the volume operator on four-valent spin network states,
which is then illustrated by explicitly calculating some
of the spectral formulae in Sec.4. In Sec.5 we summarize
and discuss our results.

\vskip2cm

\line{\ch 2 Defining the volume operator\hfil}

In order to fix the notation, we will first summarize the main
ingredients of the customary Hamiltonian lattice formulation for gauge
theories [8]. Our lattice is a cubic $N\times N\times N$-lattice, with
periodic boundary conditions, i.e. the topology of a three-torus. The
basic operators associated with each lattice link $l$ are an
$SU(2)$-link holonomy $\hat V$, together with its inverse
$\hat V^{-1}$, and a pair of canonical momentum operators $\hat p^+_i$
and $\hat p^-_i$, where $i$ is an adjoint index. The operator
$\hat p^+_i(n,\hat a)$ is based at the vertex $n$, and is
associated with the link $l$ oriented in the positive $\hat
a$-direction. By contrast, $\hat p^-_i(n+\hat 1_{\hat a},\hat a)$ is
based at the vertex displaced by one lattice unit in the
$\hat a$-direction, and associated with the inverse link
$l^{-1}(\hat a)=l(-\hat a)$. The wave functions are elements of
$\otimes_l L^2(SU(2),dg)$, with the product taken over all links, and
$dg$ is the Haar measure. The basic commutators are

$$
\eqalign{
&[\hat V_A{}^B(n,\hat a),\hat V_C{}^D(m,\hat b)]=0\cr
&[\hat  p^+_i(n,\hat a),\hat V_A{}^C(m,\hat b)]=
-\frac{i}{2}\,\d_{nm}\d_{\hat a\hat b}\, \tau_{iA}{}^B\hat V_B{}^C
(n,\hat a),\cr
&[\hat  p^-_i(n,\hat a),\hat V_A{}^C(m,\hat b)]=
-\frac{i}{2}\,\d_{nm}\d_{\hat a\hat b}\,\hat V_A{}^B
(n,\hat a) \tau_{iB}{}^C,\cr
&[\hat p^\pm_i(n,\hat a),\hat p^\pm_j(m,\hat b)]=
\pm i\, \d_{nm}\d_{\hat a\hat b}\, \e_{ijk}\, \hat p_k^\pm
(n,\hat a),\cr
&[\hat p^+_i(n,\hat a),\hat p^-_j(m,\hat b)]=0,}\eqno(2.1)
$$

\ni where $\e_{ijk}$ are the structure constants of $SU(2)$. In terms
of an explicit parametrization by four complex parameters $\a_i$,
$i=0\dots 3$, $\sum_i\a_i^2 =1$, the operators for a single link
$(n,\hat a)$ are given by

$$
\eqalign{
&\hat V_A{}^B=\left( \matrix{\a_0+i\a_1&\a_2+i\a_3,\cr
    -\a_2+i\a_3&\a_0-i\a_1}\right)=\a_0\,\one+\sum_{i=1}^3 \a_i\tau_i,\cr
&\hat p^\pm_1=\frac{i}{2}
(\a_1\del_0-\a_0\del_1\pm\a_3\del_2\mp\a_2\del_3),\cr
&\hat p^\pm_2=\frac{i}{2}
(\a_2\del_0\mp\a_3\del_1-\a_0\del_2\pm\a_1\del_3),\cr
&\hat p^\pm_3=\frac{i}{2}
(\a_3\del_0\pm\a_2\del_1\mp\a_1\del_2-\a_0\del_3),} \eqno(2.2)
$$

\ni where in the first line we have defined the three $\tau$-matrices.
In the continuum theory, the classical expression for the volume of a
spatial region ${\cal R}\subset\Sigma$ is given by

$$
{\cal V}({\cal R})=\int_{\cal R} d^3x\;\sqrt{\det g}=
\int_{\cal R} d^3x\;\sqrt{\frac{1}{3!} |\e_{abc}\,\e^{ijk} E^a_i
 E^b_j  E^c_k |},\eqno(2.3)
$$

\ni where $E^a_i$ are the dreibein variables introduced earlier
(corresponding to the generalized electric fields in a gauge
theoretic language). Taking into account the continuum limit of the
classical lattice variables,

$$
p^\pm_i(n,\hat a) \buildrel{a\rightarrow 0}\over\longrightarrow  a^2
\tilde E^a_i(n) +O(a^3),\eqno(2.4)
$$

\ni as the lattice spacing $a$ goes to zero, we define the lattice
analogue of (2.3) as

$$
{\cal V}_{\rm latt}=\sum_{n\in{\cal R}} \sqrt{\frac{1}{48}
|\e_{abc}\,\e^{ijk}\,
(p^+_i(n,\hat a) +p^-_i(n,\hat a))
(p^+_j(n,\hat b) +p^-_j(n,\hat b))
(p^+_k(n,\hat c) +p^-_k(n,\hat c)) |}.\eqno(2.5)
$$

\ni For consistency, we have averaged over the momenta of both
orientations\note{This choice does not alter the conclusions of [3],
where the unaveraged operator was used. The eigenvalues reported in [3]
are merely changed by a constant overall factor, e.g., $c=\frac14$ for
four-valent intersections.}.
The translation of this expression
to the quantum theory is not well defined a priori, because of the
presence of both the modulus and the square root. However, since
both the $\hat p^\pm$ and therefore also the operators

$$
\hat D(n):=\frac18\, \e_{abc}\,\e^{ijk}
(\hat p^+_i(n,\hat a) +\hat p^-_i(n,\hat a))
(\hat p^+_j(n,\hat b) +\hat p^-_j(n,\hat b))
(\hat p^+_k(n,\hat c) +\hat p^-_k(n,\hat c))
\eqno(2.6)
$$

\ni are selfadjoint, we may go to a Hilbert space basis of
simultaneous  eigenfunctions of all the $\hat D(n)$ and {\it define}
the operator

$$
\hat {\cal V}_{\rm latt}=\sum_n \sqrt{\frac{1}{3!}|\hat D(n) |}\eqno(2.7)
$$

\ni through the square roots of the moduli of the eigenvalues of the
$\hat D(n)$ in that basis. (Note that no operator ordering problem
occurs in the definition of $\hat D(n)$.)
As already mentioned in the introduction, the diagonalization of the
volume operator is most easily achieved starting from a set of spin
network states on the lattice. These are certain (anti-)symmetrized,
real linear combinations of Wilson loops. (A Wilson loop on the
lattice is a gauge-invariant function of the form Tr$V(l_1)V(l_2)...
V(l_k)$, where $\gamma=l_1\circ l_2\circ ...\circ l_k$ is a closed
loop of lattice links.) A spin network associates a positive
``occupation number" with each lattice link, counting the number of
(unoriented) flux lines of basic spin-$\frac12$ representations along
the link, and also keeps track of the way in which those flux lines are
contracted gauge-invariantly at the vertices (see [1] for more
details).

The operators $\hat D(n)$ have a particularly simple action on spin
networks, because they do not change their support (in terms of the
flux line numbers). Thus only finite-dimensional rearrangements occur
within each subset of states sharing the same occupation numbers, and
the diagonalization of $\hat {\cal V}_{\rm latt}$ can be
performed separately in these finite-dimensional eigenspaces [2].
In this respect, the structure of the volume operator is much simpler
than that of the Hamiltonian operator, which (at least on the lattice)
changes the support of Wilson loops [4]. Since an operator $\hat D(n)$
acts only on links adjacent to the vertex $n$, and neighbouring
$\hat D(n)$'s commute, it is sufficient to study its action on spin
networks locally around a single vertex. This will be the subject of
the next section.

\vskip2cm

\line{\ch 3 Deriving the spectrum on four-valent spin
networks\hfil}

Because of the cubic geometry of the lattice, the spin networks that
can be defined on it are at most six-valent.

\epsffile{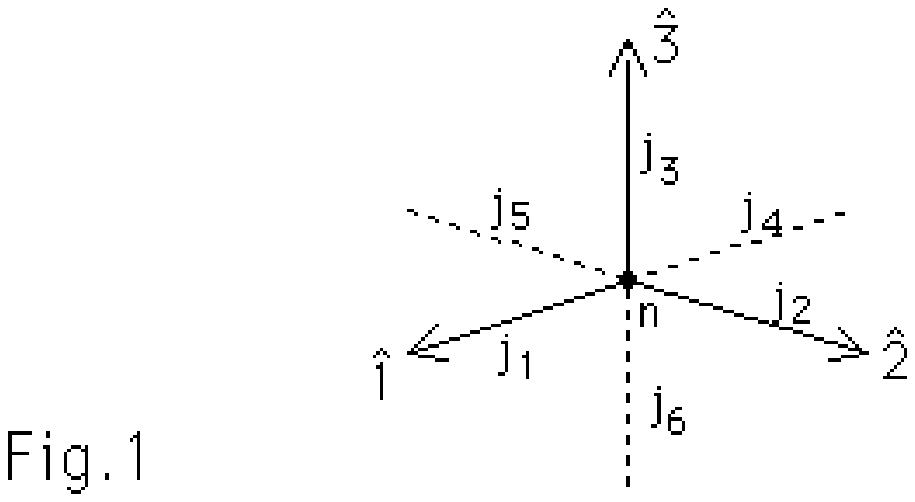}

Fig.1 illustrates our labelling of the link directions meeting at a
vertex $n$. One ingredient in describing a spin network locally around
$n$ is a 6-tuple $\vec j$ of integers
$j_i\geq 0$ giving the occupation numbers $(j_1,\dots,j_6)$ of the
links $((n,\hat 1),(n,\hat 2),(n,\hat 3);(n,-\hat 1),(n,-\hat
2),(n,-\hat 3)) \equiv ((n,\hat 1),(n,\hat 2),(n,\hat 3);
(n-\hat 1,\hat 1),(n-\hat 2,\hat 2),
(n-\hat 3,\hat 3))$ intersecting at $n$. Since the flux lines are
to be contracted at the vertex, their sum $j:=\sum_{i=1}^6 j_i$ is
an even integer. Next one has to specify how the $j$ flux lines are
joined pairwise at $n$ to ensure gauge-invariance.

By convention we may join a flux line along the positive $\hat
1$-direction, say, only to a flux line from one of the other five
links, and not from the same link (i.e. we forbid ``retracings"). This
leads to a constraint on the occupation numbers: any $j_i$ has to be
equal to or smaller than the sum of the remaining $j_k$, i.e.
$j_i\leq\sum_{k\not= i} j_k$. Given a contraction of the flux lines
at the vertex, the spin network consists of a weighted linear
combination of $j_1! j_2! j_3! j_4! j_5! j_6!$ Wilson loops
corresponding to all possible permutations of flux lines associated
with each of the six links. The weight factors are given by
$(-1)^{(P+N)}$, where $P$ is the parity of the flux line permutation
and $N$ the number of closed loops in a multiple Wilson loop that
is obtained by arbitrarily completing the local link configuration
around the vertex $n$. The relative weights of the set of multiple
Wilson loops thus obtained is independent of such an extension. To
obtain a complete spin network state, this (anti-)symmetrization of
course has to be performed around every vertex contained in the state.

Since the diagonalization of the volume operator is algebraically
rather involved, we will restrict ourselves to the simplest
non-trivial case, namely that of spin networks with at most
four-valent intersections. This is consistent since the volume
operator maps the set of such states into itself. In a previous
paper we have shown that spin networks are annihilated locally by
$\hat D(n)$ at trivalent intersections, and more generally at
intersections for which there exists only a single, unique contraction
of flux lines [3]. From our calculations on four-valent spin networks
below one recognizes this as part of a general pattern, namely
that eigenvalues occur in pairs of opposite sign.

Without loss of generality we may restrict our attention to
four-valent vertices with occupation numbers of the form
$\vec j=(p,q,r;s,0,0)$, with $p\geq q\geq r\geq s>0$. A moment's
consideration reveals that the number of spin network states one
may construct from this link configuration is $(x+1)(x+2)/2$,
where $x={\rm Min}\{s, \frac12 (-p+q+r+s)\}$. However,
these spin networks are not all linearly
independent, due to the existence of the so-called Mandelstam
constraints (see, for example, [9]). It turns out that the number of
linearly independent four-valent spin networks is $x+1$. Thus it grows
only linearly in $x$, whereas the total number of states
is proportional to
$x^2$. (It is a question of semantics whether by ``spin networks"
one means the full set of (anti-)symmetrized states as introduced
above or only an independent, already orthogonalized basis set
- we have been using it in the former sense.)

\vskip0.4cm
\epsffile{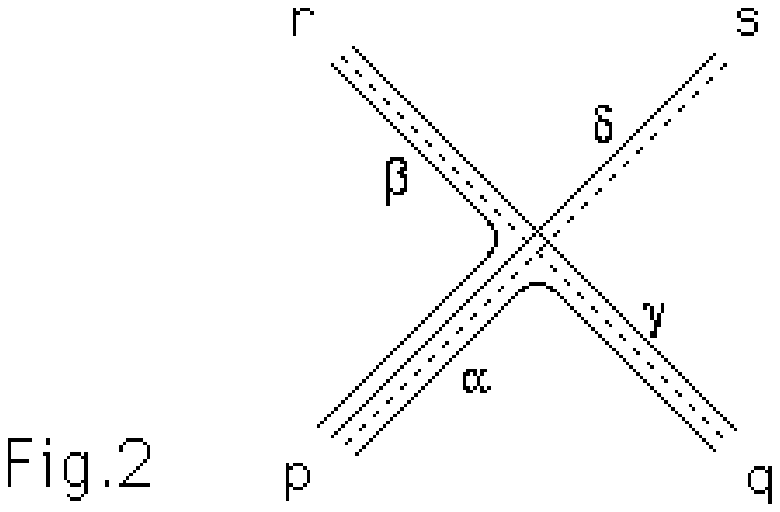}

An alternative way of parametrizing the occupation numbers is given by
four integers $(\a,\b,\g,\d)$, where $p=\a+\b+\d$, $q=\a+\g$,
$r=\b+\g$, and $s=\d$ (see Fig.2; by drawing $\a$ etc. as connected
pieces of incoming and outgoing flux lines we do not mean to indicate
that the associated spin network states will share the same routings
at the intersection; it merely gives us another way of labelling
a flux line configuration.). Note that this transformation is
invertible. The variable change is useful because of the
following construction. Abandon for the moment the restrictions $p\geq
q\geq r\geq s$, keeping however the inequalities $p\leq (q+r+s)$,
$q\leq (p+r+s)$, $r\leq (p+q+s)$, $s\leq (p+q+r)$ and the condition
$(p+q+r+s){\rm mod} \, 2=0$. Given a fixed $x\geq 1$, we define the
``fundamental link configuration" to be the one with minimal total
occupation number $(p+q+r+s)$. For this configuration we obviously have
$(p,q,r,s)=(x,x,x,x)$, or $(\a,\b,\g,\d)=(0,0,x,x)$. Now
observe that every allowed link configuration with the same $x$ can be
obtained by adding 4-tuples of non-negative integers $(\Delta \a,
\Delta \b,\Delta \g,\Delta \d)$ to $(\a,\b,\g,\d)=(0,0,x,x)$. The
converse is not true, because adding such a quadruplet may change
$x$.

Our aim is to derive general formulae for the eigenvalues of $\hat
D(n)$ on independent spin networks for fixed $x$. This can be
achieved in a series of steps. Starting from a link configuration
$(\a,\b,\g,\d)=(0,0,x,x)$, one extends it arbitrarily to obtain a
multi-loop, and then determines the corresponding set of $(x+1)(x+2)/2$
spin networks. A subset of $x+1$ linearly independent states can be
easily determined using, for example, the relations derived in [9].
Note that, due to the total (anti-)symmetry of the spin network states,
loop configurations containing retracings on any of the links meeting
at $n$ can be set identically to zero. Next one computes the action
of $\hat D(n)$ on the independent states, thus obtaining a
$(x+1)\times (x+1)$-matrix, which one then diagonalizes. It is
equivalent and much easier to compute the action of $\hat
D(n)$ on a single representative of the permutation equivalence class
of Wilson loop states that make up a given spin network, and then
check which permutation equivalence classes the resulting Wilson
loop functions lie in. In all of this, one must remember
to keep track of the
weight factors $(-1)^{(P+N)}$. A useful identity in evaluating the
action of $\hat D(n)$ is

$$
\e^{ijk}(\tau_i)_A{}^B (\tau_j)_C{}^D (\tau_k)_E{}^F=
2\, (\d_A^D \d_E^B \d_C^F -\d_A^F \d_C^B \d_E^D).\eqno(3.1)
$$

\ni All of the above steps may be implemented using algebraic computing
programs, like Mathematica. In the derivation of the eigenvalue formulae
at constant $x$ it is crucial to observe that it is sufficient to know
the action of $\hat D(n)$ on a small number of link configurations
$(\a,\b,\g,\d)$ (or rather their associated spin networks) ``close" to
the fundamental one, $(0,0,x,x)$, as we will now proceed to explain.

Let us adopt the shortcut described above for deriving the action
of $\hat D(n)$ on spin networks by their action on Wilson loop
representatives. Note that going from the link
configuration $(0,0,x,x)$ to $(1,0,x,x)$, say, at the level of
these Wilson loop representatives may be represented by adding a
single closed loop ``containing $\a$" (for an explicit example, see
Sec.4). It is then easy to
repeat the steps outlined above, and determine the matrix elements of
$\hat D(n)$. Call the matrices obtained in this way
$M(0,0,x,x)$ and $M(1,0,x,x)$ respectively. Using the explicit
action of $\hat D(n)$ on Wilson loop states, one may then
prove that any $M(\a,0,x,x)$, $\a=2,3,\dots$, can
be computed via

$$
M(\a,0,x,x)=M(0,0,x,x)+\a\,(\,
M(1,0,x,x)-M(0,0,x,x)\,).\eqno(3.2)
$$

Similar relations exist for other link configurations, and one
can derive a general formula expressing the matrix representation
of $\hat D(n)$ acting on the set of spin networks associated with
an arbitrary link configuration $(\a,\b,\g,\d)$,

$$
\eqalign{
&M_x(\a,\b,\g,\d)=
(1-\g+x)(1-\d+x)\,
(\,(1-\a)(1-\b)\,M_x(0,0,x,x) \cr
&\,+\a (1-\b)\,M_x(1,0,x,x)
+(1-\a)\b \,M_x(0,1,x,x)
+\a\b \,M_x(1,1,x,x)\,) \cr
&\,+(\,(\g-x) (1-\d+x)+(1-\g+x) (\d-x)\,)\,
(\,(1-\a)(1-\b)\,M_x(0,0,x+1,x) \cr
&\, +\a (1-\b) \, M_x(1,0,x+1,x)+
(1-\a)\b \,M_x(0,1,x+1,x)+
\a\b \,M_x(1,1,x+1,x)\,) }\eqno(3.3)
$$

\ni where we have introduced a subscript $x$ to indicate that the
matrix elements of $M$ refer to sets of spin networks with fixed
$x$. Formula (3.2) is of course a special case of (3.3). In the
derivation of (3.3) we have taken into account that
$M_x(0,0,x+1,x)=M_x(0,0,x,x+1)$, $M_x(1,0,x+1,x)=M_x(1,0,x,x+1)$ etc.

Given this expression, it is straightforward to
establish the explicit eigenvalue formulae for a given $x$.
To illustrate our method, we will in the next section discuss
the case $x=1$ in some detail, and also give the eigenvalue
formulae for $x=2$ and $3$.

\vskip2cm

\line{\ch 4 Computation of the spectrum for small x\hfil}

Recall first that, for $x=0$, following [3], all eigenvalues of the
volume operator vanish identically. Let us therefore turn to
the case $x=1$. The fundamental link configuration is given
by $(\a,\b,\g,\d)=(0,0,1,1)$ or, equivalently,
$(p,q,r,s)=(1,1,1,1)$, Fig.3a.

\vskip0.5cm
\epsffile{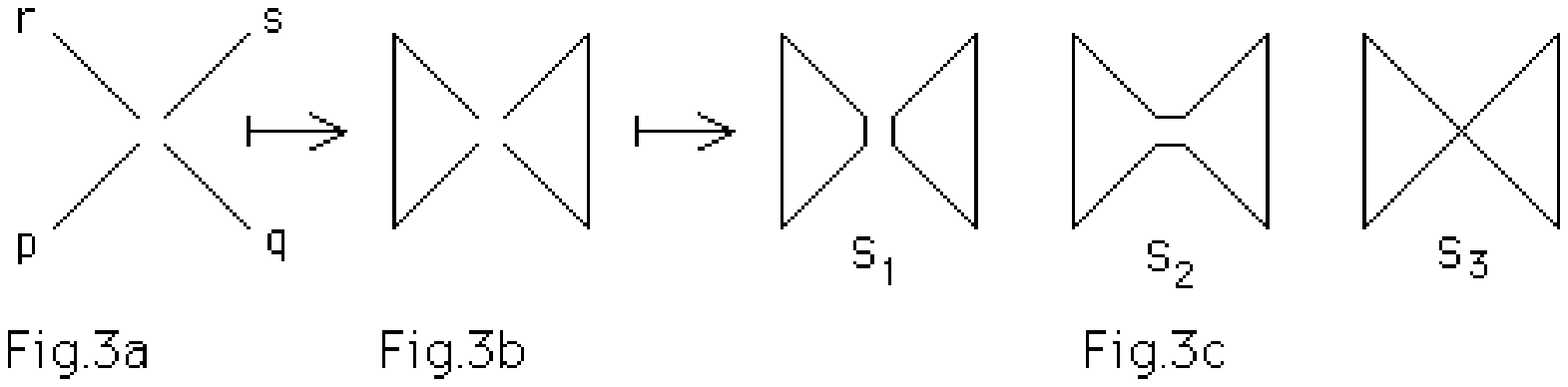}

Now close up the external ends of the four links in some way,
as shown schematically in
Fig.3b. There are then three spin networks, $S_1$, $S_2$
and $S_3$, obtained by connecting the internal ends of the links
meeting at the vertex $n$ in various ways (Fig.3c). Since the links
are singly occupied, no permutations have to be taken into account.
The Mandelstam identity for this set of spin networks is
$S_1-S_2-S_3=0$. We choose as an independent set the states
$S_1$ and $S_2$.
One then computes the matrix representation for $\hat D(n)$ on
these states, which is given by $M_1(0,0,1,1)=\frac{3 i}{8}\left(
\matrix{1&-2\cr 2&-1}\right)$. According to our reasoning of the
previous section, we have to compute another seven matrices $M_1$.

Consider next the link configuration $(\a,\b,\g,\d)=(1,0,1,1)$.
This means adding two links to the previous configuration, as
illustrated
in Fig.4a and Fig.4b. The analogue of Fig.3c is shown in Fig.4c.

\epsffile{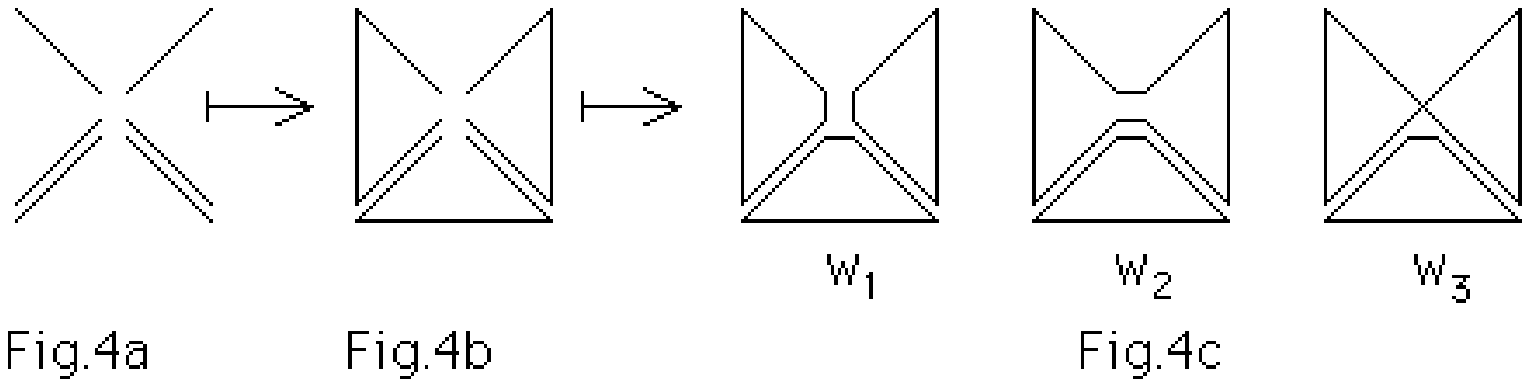}

The three configurations $W_1$, $W_2$ and $W_3$ are now not spin
networks themselves (because no (anti-)symmetrization has been
performed), but can be considered as Wilson loop representatives of
permutation equivalence classes, as explained earlier. For
the corresponding spin networks $S_1'$, $S_2'$ and $S_3'$, by
construction $S_1'-S_2'-S_3'=0$ continues to hold, and one computes
$M_1(1,0,1,1)=\frac{3 i}{8}\left(\matrix{2&-3\cr 4&-2}\right)$ in
the basis $\{S_1',S_2'\}$. Proceeding similarly with the other
relevant link configurations, one obtains for (3.3)

$$
M_1(\a,\b,\g,\d)=\frac{3 i}{8}
\left(\matrix{(1+\a)(1+\b) & -(1+\b)(\a+\g+\d)\cr (1+\a)(\b+\g+\d)&
-(1+\a)(1+\b)}\right),
\eqno(4.1)
$$

\ni from which one computes the eigenvalues of $\hat D(n)$ on
link configurations $(\a,\b,\g,\d)$ as

$$
\pm\frac{3}{8} \sqrt{(1+\g+\d) (1+\a)(1+\b)(1+\a+\b+\g+\d)}.\eqno(4.2)
$$

\ni Note that we have used the identity $(\g-1)(\d-1)=0$, valid
for $x=1$. Going back to the notation $(p,q,r,s)$
for link configurations, with $p\geq q\geq r\geq s >0$,
one finds for the first few configurations
the following eigenvalues:

\vskip1cm
\vbox{\offinterlineskip
 \halign{&\vrule#&\strut\quad\hfil#\quad\cr
 height2pt&\omit&&\omit&&\omit&&\omit&&\omit&&\omit&&\omit&\cr
 &(p,q,r,s)\hfil&&(1,1,1,1)&&(2,2,1,1)&&(3,2,2,1)&&(3,3,1,1)&&
 (3,3,3,1)&&(4,2,2,2)&\cr
 height2pt&\omit&&\omit&&\omit&&\omit&&\omit&&\omit&&\omit&\cr
 \noalign{\hrule}
 height2pt&\omit&&\omit&&\omit&&\omit&&\omit&&\omit&&\omit&\cr
 &eigenvalues&&$\pm\frac38 \sqrt{3}$\hfil&&$\pm\frac{3}{2\sqrt{2}}$
 \hfil&&$\pm\frac{3\sqrt{5}}{4}$
 \hfil&&$\pm\frac38 \sqrt{15}$\hfil&&$\pm\frac{3\sqrt{3}}{2}$
\hfil&&$\pm\frac{3\sqrt{3}}{2}$
 \hfil&\cr
 height2pt&\omit&&\omit&&\omit&&\omit&&\omit&&\omit&&\omit&\cr}}

Similarly, we have computed the eigenvalues of $\hat D(n)$ for
$x=2$. For each link configuration there are six spin network states,
out of which three are linearly independent. For the matrix
$M_2(\a,\b,\g,\d)$, there is one
zero-eigenvalue and a pair of non-vanishing ones,

$$
\pm\frac{3}{4\sqrt{2}}\sqrt{-(2+\a)(2+\b)(2+\a+\b)+
(\g+\d-2)(\a+\b+\g+\d)(2 \a\b+3\a+3\b+4)}. \eqno(4.3)
$$

In principle, formula (3.3) can be used also for the unreduced matrices,
i.e. before the linearly dependent spin networks are eliminated.
However, it is unfortunately not true that such states
automatically have vanishing eigenvalues. Of course, if complex
eigenvalues occur it is clear that they must correspond to spin
networks that vanish modulo the Mandelstam constraints. But even
non-zero, real eigenvalues occur for these spurious eigenvectors.
(This happens, for example, in the case $x=3$.)
Again, we give a list of the first few eigenvalues corresponding to
the matrix $M_2(\a,\b,\g,\d)$:

\vskip1cm
\vbox{\offinterlineskip
 \halign{&\vrule#&\strut\quad\hfil#\quad\cr
 height2pt&\omit&&\omit&&\omit&&\omit&&\omit&&\omit&&\omit&\cr
 &(p,q,r,s)\hfil&&(2,2,2,2)&&(3,3,2,2)&&(4,3,3,2)&&(4,4,2,2)&&
 (4,4,4,2)&&(5,3,3,3)&\cr
 height2pt&\omit&&\omit&&\omit&&\omit&&\omit&&\omit&&\omit&\cr
 \noalign{\hrule}
 height2pt&\omit&&\omit&&\omit&&\omit&&\omit&&\omit&&\omit&\cr
 &eigenvalues&&$\pm\frac{3\sqrt{3}}{2}$\hfil&&
 $\pm\frac{3\sqrt{13}}{2\sqrt{2}}$\hfil&&
 $\pm\frac{9\sqrt{3}}{2\sqrt{2}}$
 \hfil&&$\pm\frac32\sqrt{11}$\hfil&&
 $\pm\frac92\sqrt{3}$\hfil&&$\pm\frac92\sqrt{3}$
 \hfil&\cr
 height2pt&\omit&&\omit&&\omit&&\omit&&\omit&&\omit&&\omit&\cr}}

For higher $x$, the algebra becomes progressively more complicated,
but no problems occur in principle in determining the eigenvalues.
Here are the eigenvalues of the matrix $M_3(\a,\b,\g,\d)$ for the
case $x=3$, which come in two pairs of opposite sign, namely,

$$
\pm\frac38 \,\sqrt{I_1\pm\sqrt{I_2}},\eqno(4.4)
$$

\ni where

$$
\eqalign{
&I_1=5 \mu(\a+2)(\b+2)\nu-3\mu (\mu+2)+3 (\a+\b+6)(\g+\d)+
 3(\a\b-11)\cr
&I_2=16 (\mu+2)^2 \nu^2 (\a+1)(\b+1)(\a+3)(\b+3)+
     5 (\mu+2)^2 \nu^2 (5 \a^2+5 \b^2 +8\a +8\b-6\a\b) \cr
&\,+ 64 (\mu+2)^2 \nu^2 -
     64 (\mu+2)\nu^2 (\a+2)^2 (\b+2)^2 \cr
&\,+ 12 (\mu+2)^3 ( 10 (\a+2)(\b+2)-3(\a+2)^2 -3(\b+2)^2) )\cr
&\,+ 4 (13\mu +15) \nu(\a+2)^2 (\b+2)^2
   - 6 \nu^2 (\a+2)(\b+2)(5 \a^2 +5\b^2 +48\a +48\b+244)\cr
&\,+ 36 \nu^2 (2 \a^3 +17 \a^2 +44\a +2\b^3 +17\b^2 +44\b+72)
   -94 \nu(\a+2)^2 (\b+2)^2 (\a+\b+10)\cr
&\,-6 \nu (\a+2)(\b+2)(5 \a^3 +54 \a^2 +316 \a+5 \b^3 +
       54 \b^2 +316 \b +1152)\cr
&\, +108 \nu( (\a+2)^2 (a+4)^2 +(\b+2)^2 (\b+4)^2 )
    +36 (\a+4)^3 (\a+2)^2 +36 (\b+4)^3 (\b+2)^2 \cr
&\, +18 (\a+4)(\b+4)(\a+2)^2 (\b+2)^2
    -960 (\a+2)(\b+2) \cr
&\, +9 (\a+2)(\b+2)(\a+4)(\b+4)(\a^2-26\a +\b^2 -26\b -112)\cr
&\, +6 (\a+2)(\b+2) ((\a+2)(-5 \a^2+16\a+64) +
       (\b+2)(-5 \b^2+16\b+64)) ,}\eqno(4.5)
$$

\ni with the abbreviations $\mu=\a+\b+\g+\d$ and $\nu=\g+\d-4$.
For small occupation numbers $(p,q,r,s)$, some explicit eigenvalues
are given by:

\vskip1cm
\vbox{\offinterlineskip
 \halign{&\vrule#&\strut\quad\hfil#\quad\cr
 height2pt&\omit&&\omit&&\omit&&\omit&&\omit&\cr
 &(p,q,r,s)\hfil&&(3,3,3,3)\hfil&&(4,4,3,3)\hfil&&(5,4,4,3)
 \hfil&&(5,5,5,3)\hfil&\cr
 height2pt&\omit&&\omit&&\omit&&\omit&&\omit&\cr
 \noalign{\hrule}
 height2pt&\omit&&\omit&&\omit&&\omit&&\omit&\cr
 &eigenvalues&&$\pm\frac98\sqrt{19\pm 16}$\hfil&&
 $\pm\frac94 \sqrt{9\pm\sqrt{57}}$\hfil&&
 $\pm\frac{9}{4\sqrt{2}}\sqrt{33\pm\sqrt{753}}$
 \hfil&&
 $\pm \frac{3}{\sqrt{2}}\sqrt{33\pm 27}$
 \hfil&\cr
 height2pt&\omit&&\omit&&\omit&&\omit&&\omit&\cr}}

\vskip2cm

\line{\ch 5 Summary and discussion\hfil}

We have derived a general formula for the representation matrix of
the operator $\hat D(n)$ on sets of four-valent
spin network states corresponding
to a link configuration $(p,q,r,s)$, with fixed $x$, where
$x={\rm Min}\{s, \frac12 (-p+q+r+s)\}$. This expression, formula (3.3),
is given in terms of eight matrices whose entries have to be
computed for each $x$. The spin network states are not independent,
but obey certain linear relations, the Mandelstam constraints.
The linearly dependent states
have to be identified explicitly, otherwise one obtains spurious
eigenvalues after diagonalization. At this moment, the only
restriction on obtaining the full spectrum on four-valent spin
networks is computing capacity. It may be possible that the
formulae obtained above for $x=1,2,3$ can be written in a more
symmetric form that can be generalized to arbitrary $x$; so far
we have not been able to do this.

We have observed that eigenvalues occur in pairs of opposite sign.
Thus, for odd $x+1$ there is a zero-eigenvalue for an eigenstate
that does not vanish modulo the Mandelstam constraints.
This means that there exist four-valent spin networks ``without volume"
(which by construction are non-planar; the planar ones are all
annihilated by $\hat {\cal V}$ because of antisymmetry).
Another consequence is that after taking the modulus, as is
necessary for constructing the local volume operator $\hat {\cal V}(n)$,
all non-zero eigenvalues are (at least) two-fold degenerate.

The generalization of our results to spin networks of higher
valence is algebraically more complicated; in this case already the
counting of spin networks is less straightforward. Still,
there are no obvious obstructions to deriving analogues of
our matrix formula (3.3).

As for the geometric interpretation of our eigenvalue expressions,
we have seen that their dependence on the occupation numbers
$(p,q,r,s)$ is not particularly simple, for example, they do not
just depend on the total number of links, $(p+q+r+s)$. One
possibility is to try to extract from them a
certain asymptotic behaviour,
say, as the total number of links becomes very large. For
example, for fixed $x$, one may look at spin networks corresponding
to link configurations of the form $(p,p,p,x)$, where $p$ runs
through all odd or even integers, depending on whether $x$ is odd
or even. For the cases studied in the last section, one finds
the following asymptotic behaviour of the eigenvalues of $\hat D(n)$
for large $p$ and to highest order in $p$:

$$
\eqalign{
x=1:&\quad \sim \pm \frac{3\sqrt{3}}{2^5}\, p^2\cr
x=2:&\quad \sim \pm \frac{3\sqrt{3}}{2^4}\, p^2\cr
x=3:&\quad \sim \pm \frac{3\sqrt{3}}{2^5} (2\pm 1)\, p^2}\eqno(5.1)
$$

\ni The ``local volume", $\hat {\cal V}(n)$, therefore increases
linearly in $p$, and can become arbitrarily large for the classes
of spin networks considered here. Such asymptotic relations may
be useful in estimating the contributions of certain sectors of the
Hilbert space of spin network states in numerical approximations.

\ni{\it Acknowledgement.} I am grateful to the members of the
Max-Planck-Institut for their kind hospitality.

\vskip2cm
\vfill\eject

\line{\ch References\hfil}

\item{[1]} Baez, J.B.: Spin network states in gauge theory, to
  appear in {\it Adv. Math.}, e-Print Archive: gr-qc 9411007;
  Spin networks and nonperturbative quantum gravity, to appear in the
  proceedings of the AMS Short Course on Knots and Physics,
  e-Print Archive: gr-qc 9504036

\item{[2]} Rovelli, C. and Smolin, L.: Discreteness of area and
  volume in quantum gravity, {\it Nucl. Phys.} B442 (1995)
  593-619

\item{[3]} Loll, R.: The volume operator in discretized quantum
  gravity, {\it Phys. Rev. Lett.} 75 (1995) 3048-51

\item{[4]} Loll, R.: Non-perturbative solutions for lattice quantum
  gravity, {\it Nucl. Phys.} B444 (1995) 619-39

\item{[5]} L. Smolin, personal communication

\item{[6]} A. Ashtekar, personal communication; see also Ashtekar,
  A.: Polymer geometry at Planck scale and quantum Einstein
  equations, {\it preprint} Penn State Univ., 1995

\item{[7]} Barbero, J.F.: Reality conditions and Ashtekar variables:
  A different perspective, {\it Phys. Rev.} D51 (1995) 5498-506;
  Real Ashtekar variables for Lorentzian signature space-times,
  {\it Phys. Rev.} D51 (1995) 5507-10

\item{[8]} Kogut, J. and Susskind, L.: Hamiltonian formulation of
  Wilson's lattice gauge theories, {\it Phys. Rev.} D11 (1975)
  395-408; Kogut, J.B.: The lattice gauge theory approach
  to quantum chromodynamics, {\it Rev. Mod. Phys.} 55 (1983) 775-836

\item{[9]} Loll, R.: Independent SU(2)-variables and the reduced
  configuration space of SU(2)-lattice gauge theory, {\it Nucl.
  Phys.} B368 (1992) 121-42; Yang-Mills theory without
  Mandelstam constraints, {\it Nucl. Phys.} B400 (1993) 126-44

\end